\documentclass[aps,prl,twocolumn,10pt,superscriptaddress]{revtex4-2}

\usepackage{amssymb,amsfonts,amsmath}
\usepackage{times}
\usepackage{graphicx}
\usepackage{color}
\usepackage{float}
\usepackage{cancel}
\usepackage[T1]{fontenc}
\usepackage{lipsum}
\usepackage[latin9]{inputenc}
\usepackage{babel}
\usepackage{bbold}
\usepackage{wasysym}
\usepackage{multirow}
\usepackage[dvipsnames]{xcolor}
\usepackage{pifont}
\usepackage{microtype}
\usepackage{physics}
\usepackage{xcolor}

\usepackage[linktocpage=true,
  colorlinks=true, 
  pdfborder={0 0 0},
  linkcolor=blue,
  citecolor=red,
  filecolor=yellow,
  urlcolor=blue,
  bookmarks,
  pdfauthor={},
]{hyperref}

\newlength{\figurewidth}
\newlength{\pagewidth}
\usepackage{longtable}
\newcommand{\tc}{$T_\mathrm{c}$}
\newcommand{\hf}{H$_\text{f}$}
\newcommand{\hdos}{H$_\text{DOS}$}

\begin{document}

\title{Refining \tc\ Prediction in Hydrides via Symbolic-Regression-Enhanced Electron-Localization-Function-Based Descriptors}

\author{Francesco Belli}\affiliation{Department of Chemistry, State University of New York at Buffalo, Buffalo, NY 14260-3000, USA}
\author{Sean  Torres}\affiliation{Department of Chemistry, State University of New York at Buffalo, Buffalo, NY 14260-3000, USA}
\author{Julia Contreras-Garc\'ia}\affiliation{Laboratoire de Chimie Th\'eorique (UPMC) CC137, Paris 75005, France}
\author{Eva Zurek}\email{ezurek@buffalo.edu}\affiliation{Department of Chemistry, State University of New York at Buffalo, Buffalo, NY 14260-3000, USA}


\begin{abstract}
Hydrogen-based materials are able to possess extremely high superconducting critical temperatures, \tc s, due to hydrogen's low atomic mass and strong electron-phonon interaction. Recently, a descriptor based on the Electron Localization Function (ELF) has enabled the rapid estimation of the \tc\ of hydrogen-containing compounds from electronic networking properties, but its applicability has been limited by the small size and homogeneity of the training dataset used. Herein, the model is re-examined compiling a publicly available combined dataset of 244 binary and ternary hydride superconductors. Our analysis shows that though ELF-based networking remains a valuable descriptor, its predictive power declines with increasing compositional complexity. However, by introducing the molecularity index, defined as the highest value of the ELF at which two hydrogen atoms connect, and applying symbolic regression, the accuracy of the predictions can be substantially enhanced. These results establish a more robust framework for assessing superconductivity in hydride materials, facilitating accelerated screening of novel candidates through integration with crystal structure prediction methods or high-throughput searches.

\end{abstract}

\maketitle

\section{Introduction}
Hydrogen-based compounds are currently among the most extensively studied superconductors due to their exceptionally high predicted and measured superconducting critical temperatures (\tc s)~\cite{Zurek:2021k,Zurek:2022k,10.1093/nsr/nwad270,Zhao:2023b,Pickard:2020a,Boebinger:2025a}. Their remarkable behavior arises from the light mass of the hydrogen atoms, resulting in high vibrational frequencies, and their lack of core electrons, which facilitates strong electron-phonon interactions~\cite{Ashcroft:2004a,Ashcroft:1968a}. In most cases, pressure is required to stabilize the unique chemical compositions and hydrogenic motifs -- key for metallization and concomitant superconductivity -- in these systems. Notable examples of synthesized phases and their \tc s include: H$_3$S (203~K, 200~GPa) ~\cite{Drozdov:2015a}, LaH$_{10}$ (250~K, 150~GPa) ~\cite{LaH10_Somayazulu,Drozdov2019superconductivity}, YH$_9$ (243~K, 201~GPa)~\cite{kongSuperconductivity243YttriumHydrogen2021}, YH$_6$ (224~K, 166~GPa)~\cite{troyan2021anomalous}, CaH$_6$ (215~K, 172~GPa)~\cite{PhysRevLett.128.167001}, and, more recently, ternary hydrides such as (La,Be)H$_8$ (120~K, 80~GPa)~\cite{PhysRevLett.130.266001}, (La,Y)H$_{10}$ (253~K, 183~GPa)~\cite{SEMENOK202118}, (La,Ce)H$_9$ (148~K, 97~GPa)~\cite{bi2022giant}, and (La,Al)H$_{10}$ (223~K, 164~GPa)~\cite{10.1093/nsr/nwad107}. Experiments have reported the superconducting gap structure in H$_3$S, confirming the phonon-mediated mechanism of pairing~\cite{DuFeng:2025a}, and local magnetometry experiments provided evidence for the Meissner effect in CeH$_9$~\cite{Bhattacharyya:2024a}. 

Computational methods have played a pivotal role in the discovery of the high-pressure hydrides by pinpointing targets for synthesis and aiding experimental characterization~\cite{Zurek:2018m,Livas,Zurek:2014i}. Moreover, computational studies have been key in providing insight on the microscopic mechanism of superconductivity in this class of compounds~\cite{errea2020quantum,doi:10.1073/pnas.1704505114}. The theoretical framework for conventional superconductors is highly developed~\cite{PhysRevB.87.024505,kogler2025isome}, allowing for accurate predictions of superconducting properties with remarkable precision~\cite{PhysRevB.106.L180501,PhysRevB.110.L140502,PhysRevB.108.214523} (though for high-pressure hydrides most \tc\ estimates have come with large error bars, in part due to their relatively large unit cells). Moreover, crystal structure prediction (CSP) techniques have resulted in an unprecedented proliferation of proposed hydrogen-rich compounds, many of which are predicted to be promising candidates for high-\tc\ superconductivity, even at moderate pressures~\cite{PhysRevLett.123.097001,Zurek:2024d,Zurek:2024f,duan2014pressure,PhysRevB.91.180502,doi:10.1021/acs.jpcc.5b12009,PhysRevLett.119.107001,doi:10.1073/pnas.1704505114,doi:10.1021/jacs.3c14205}. 

The high temperature superconductivity in the high-pressure hydrides stems from the weakening of the intramolecular H-H bonds and formation of extended networks of weakly bonded hydrogen atoms under pressure~\cite{belli2021strong}. For this reason high-pressure clathrate hydrides currently hold the record for the highest \tc\ values~\cite{10.1093/nsr/nwad270}, and ambient pressure transition metal hydrides, where hydrogen is confined within interstitial sites in the metallic sublattices, are computed~\cite{belli2024efficient,D5MH00177C,PhysRevLett.111.177002} and measured~\cite{stritzker1974high,PhysRevB.10.3818} to exhibit significantly lower \tc s. Detailed theoretical analysis suggests that the likelihood of synthesizing a hydride-based room-temperature superconductor at ambient pressure is miniscule~\cite{gao2025maximum}. However, perovskite-type structures~\cite{sanna2024prediction,PhysRevLett.132.166001,dangic2024ambient,Zurek:2024i}, and XMH$_8$ stoichiometry compounds (where X and M represent non hydrogen atoms) that are related to the binary high-pressure hydrides~\cite{PhysRevB.104.L020511,PhysRevB.106.134509,PhysRevB.104.L020511,lucrezi2022silico,Zurek:2022f} have emerged as promising candidates for reasonably high-\tc\ superconductivity at pressures approaching ambient.

Despite the successes of CSP in identifying a large number of hydride superconductors, some of which have been synthesized, CSP-based approaches are fraught with numerous challenges. First, because CSP algorithms are metaheuristics, one can never be certain that the ground state has been found. Secondly, the choice of the DFT functional, neglect of finite temperature contributions to the free energy as well as quantum nuclear effects and anharmonicity, and the inability to fully sample the compositional space are approximations that further hinder CSP~\cite{Zurek:2018m}. Finally, it is not clear if the synthetic techniques access metastable phases and if the metastable systems that are computationally predicted are kinetically (thermally) stable at finite temperatures. Finally, because DFT calculations of the electron-phonon properties are computationally expensive, in the past they have only been performed for the most promising high-symmetry systems. As a result, it is possible that high-\tc\ phases with large unit cells, vacancies, or potential disorder have been overlooked. 

To mitigate the large cost associated with \emph{ab initio} CSP searches, approaches that employ bespoke machine learning (ML) interatomic potentials for filtering the initial set of structures that are identified, followed by DFT relaxation of the most promising species, have been put forward~\cite{salzbrenner2023developments,pickard2022ephemeral}. ML generative models whose training set included high-pressure phases have been developed~\cite{luo2024deep}, and data-driven techniques~\cite{jiang2024data}, as well as high-throughput screenings that focus on structural templates~\cite{lucrezi2022silico,GENG2024101443}, have been employed. Additional efforts have focused on estimating the critical temperature through ML-assisted methods or easily computable properties~\cite{gibson2025accelerating}, though models that estimate \tc\ specifically for hydrides are scarce~\cite{jiang2024data,MA2023101233,PhysRevB.101.144505,belli2021strong,PhysRevB.110.174515,D4SC04465G}. Many of these techniques are based on observed correlations of the \tc\ with simple descriptors such as electronegativity, atomic mass, composition, space group, volume, density of states, and interatomic distances~\cite{PhysRevB.101.144505,belli2021strong,jiang2024data,wrona2025high,Saha:2023a}, and, notably, it has been shown that the magnitude of \tc\ in hydrides is closely linked to the ability of hydrogen atoms to form extended electronic networks within the crystal structure~\cite{belli2021strong,PhysRevB.110.174515,D4SC04465G}.

This last discovery is particularly interesting as it allows for a quantitative assessment of \tc\ through electronic networking properties estimated by the electron localization function (ELF)~\cite{Becke1990}. This method, initially parametrized for binary hydride superconductors, was shown to estimate the \tc\ with an accuracy of 60~K and was applicable across both low and high \tc\ systems for the hydride superconductors known at the time~\cite{belli2021strong}. While this error may initially appear large, it needs to be stressed that such a \tc\ estimate requires only knowledge of the atomic composition, the Density of States (DOS) at the Fermi energy, and the ELF iso-surface. Further, the training data used was obtained from the literature, and for high \tc\ systems, the choice of the computational method used to estimate \tc, as well as the computational parameters (Coulomb repulsion parameter, mesh size, pseudopotential and energy-cut off), can easily result in predictions that differ by $\sim$50~K. In a follow-up study, this method was further refined to capture variations in \tc\ arising from subtle changes in composition by accounting for the presence of hydrogen molecules instead of only relying on the electronic networking properties based on the ELF~\cite{PhysRevB.110.174515}. Moreover, an automated tool for estimating \tc\ has recently been made available~\cite{D4SC04465G}, enabling seamless integration with CSP software or high-throughput workflows.

However, even this improved ELF-based \tc\ model for hydrides is limited, as the dataset used to construct it contained only 129 binaries and 21 ternaries~\cite{PhysRevB.110.174515}. By now, DFT-based CSP has exhausted the high-pressure binary hydride space, and exploration of the vast number of possible chemical combinations of ternary and quaternary hydrides has just begun~\cite{Zurek:2020k}. Moreover, experiments have reported promising superconducting properties in complex systems such as medium- and high-entropy superhydrides~\cite{Zhao:2023b}, motivating future studies of multi-component hydrides.

Herein, we re-examine the aforementioned \tc\ models based on ELF networking properties utilizing an expanded dataset of 119 binary and 125 ternary hydride superconductors. While the \tc s of compounds comprising this extended dataset were still found to be related to the crystal networking properties, confirming its utility as a predictor, the error using the previously published fits \cite{belli2021strong,PhysRevB.110.174515} was found to be significant. The compiled dataset has also been made publicly available.

We show that ELF-based fits can be improved by making use of the molecularity index~\cite{PhysRevB.110.174515}, and leveraging symbolic regression~\cite{PhysRevMaterials.2.083802}. The new fit we propose results in a more accurate estimation of the \tc\ for both binary and multinary hydride superconductors, improving predictive capability. We suggest this new fit could be a useful tool to quickly determine if a material could be a promising hydride superconductor before performing expensive first-principles calculations, or when such calculations are not possible due to large unit cell sizes. We expect this fit will become useful in high-throughput screenings and in multi-objective CSP searches~\cite{Zurek:2024e}.

\section{results}

\subsection{Systems overview}
The first study that employed the results of ELF calculations for \tc\ predictions on hydrides, and upon which this work is based, introduced a classification scheme for the type of bonding present, focusing on the crucial hydrogen atoms~\cite{belli2021strong}. Six different categories were identified: any system that contained at least one H$_2$ molecule per unit cell was classified as \textit{molecular}. If any of the H-H bonds present were weakened and elongated relative to a standard homonuclear bond, the hydrides were classified as \textit{weakly hydrogen-bonded}; examples include H$_n^{\delta\pm}$ ($n>2$) molecular units, or 1D, 2D, and 3D extended lattices. \textit{Covalent systems} were characterized by an electron-pair-sharing bond between hydrogen and another (typically $p$-block) element, while \textit{ionic systems} possessed H$^-$ units with Bader charges of at least -0.5$e$, and \textit{electride systems} contained localized pockets of charge in interstitial regions without the presence of direct hydrogen-hydrogen bonds. Finally, \textit{isolated systems} were those that did not have ELF isosurfaces above 0.25. In some cases systems could possess more than one such motif, and for these fuzzy cases a priority in the classes was established.

\begin{figure*}

\includegraphics[width=\textwidth]{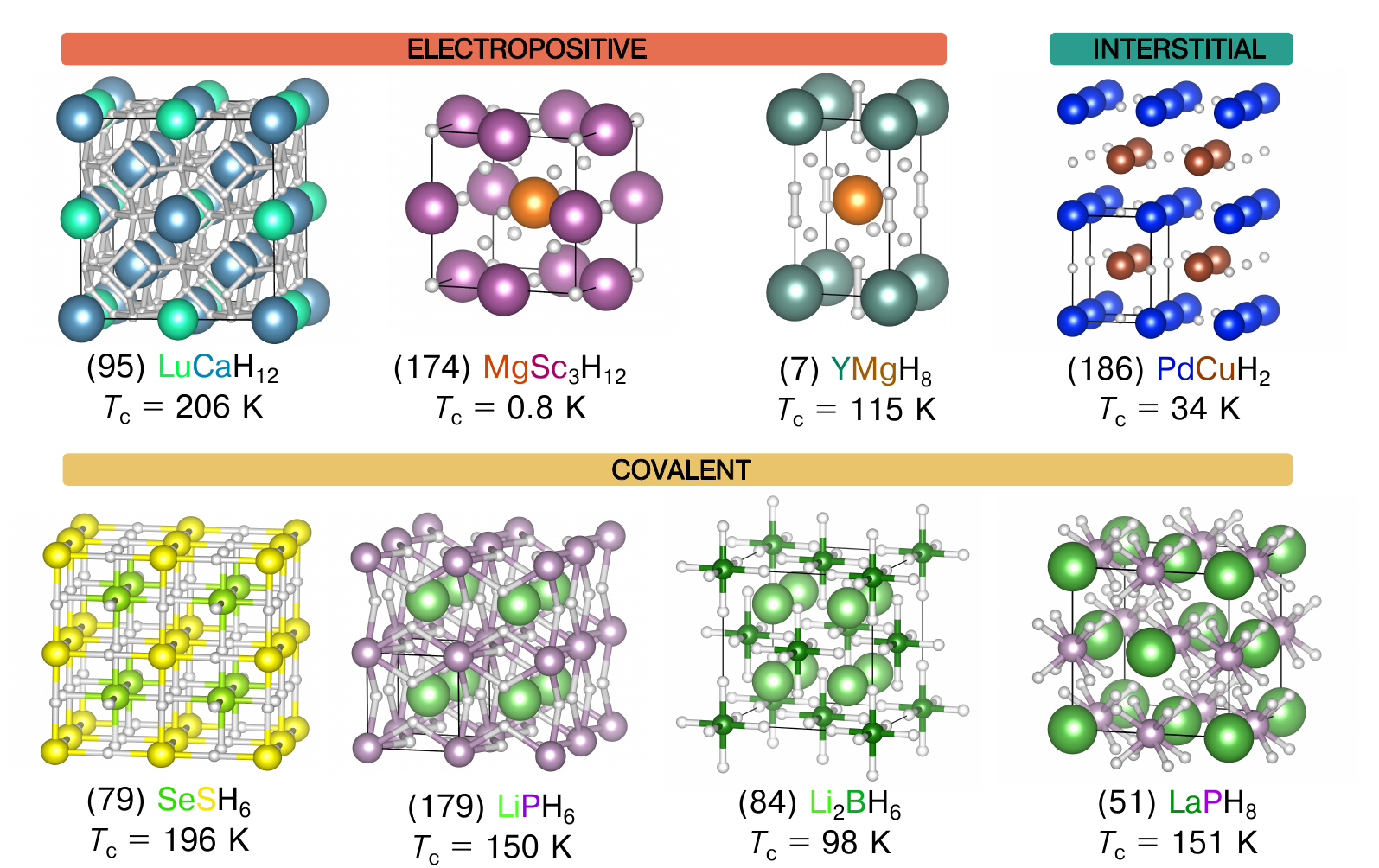}

 \caption{\label{FIG:structures}Example of the most structurally dissimilar compounds present in the dataset; the index of the structure in the dataset is given in parentheses. The \tc\ shown was obtained from the literature. Compounds 95, 174, and 7 belong to the \emph{electropositive} class. Structure 186 is an example of \emph{interstitial} hydride with a low \tc. Finally, structures 79, 179, 84 and 51 represent \emph{covalent} hydrides where the strongest interactions are mainly between hydrogen and p-block elements.}
\end{figure*}

Systems with weakly bonded hydrogenic lattices, formed by the dissociation of hydrogen and enabled by the presence of precompressor atoms, and those that contained directional bonds between hydrogen and non hydrogen (labelled as X) host atoms whose p shells were valence, were found to be good candidates for high critical temperatures. On the other hand, electrides as well as ionic and isolated systems exhibited {\tc}s lower than 50~K. All of these classes featured distinct \tc\ trends, well described by the proposed ELF-based model, and their {\tc}s were reasonably well predicted. However, the model was not without its pitfalls. For example, it was unable to resolve subtle structural differences that can significantly impact the \tc ~\cite{belli2021strong}. A clear example is the predicted compounds Li$_{2}$ScH$_{16}$ and Li$_{2}$ScH$_{17}$, which have very similar weakly bonded hydrogenic lattices but possess very different \tc\ values: 281~K and 94~K, respectively~\cite{sun2022high}. Due to their structural resemblance, the fit predicted comparable \tc\ values for both. 

To address this limitation, Di Mauro et al.~\cite{PhysRevB.110.174515} suggested the introduction of a  \emph{molecularity index} as a characterization variable for the molecularization effects leading to such \tc\ differences. However, the dataset they employed was mostly comprised of binary hydrides, containing only a limited number of ternary systems. This raises questions about the transferability of the model toward more complex hydride materials, especially as the field increasingly focuses on multinary compounds, such as ternary and quaternary hydrides, and medium-entropy hydrides. At present, no comprehensive analysis exists, including sufficient examples from this broader class of systems, that could validate the method's accuracy. 

In this study, we develop a more comprehensive ternary dataset including the previously studied binary hydrides~\cite{belli2021strong}, with an additional 125 ternary structures retrieved from the literature~\cite{PhysRevB.107.134509,https://doi.org/10.1002/adts.202100364,PhysRevB.104.L020511,PhysRevB.106.134509,PhysRevB.105.104511,PhysRevB.105.224107,PhysRevB.102.184103,PhysRevB.101.134504,PhysRevB.102.014516,PhysRevB.106.014521,PhysRevLett.128.047001,PhysRevB.106.024519,PhysRevB.100.184502,PhysRevB.104.224510,PhysRevB.104.134501,PhysRevMaterials.1.074803,doi:10.34133/2022/9784309,Tian_2015,Sun_2022,D2TC04029H,D2CP00059H,D2TC02842E,shi2021prediction,D1CP03896F,D1CP04963A,D1CP02781F,doi:10.1021/acs.jpclett.9b03856,D2CP05850B,tsuppayakorn2022stabilizing,doi:10.1021/acs.jpcc.1c08743,SAHOO2023111193,shao2019ternary,https://doi.org/10.1002/pssr.202300043,https://doi.org/10.1002/er.8705,PhysRevB.96.144518,C7CP05267G,doi:10.1021/acs.jpcc.0c09447,10.1063/5.0076728}. While binary hydrides were primarily discovered via CSP techniques, over time clear structural patterns associated with promising superconducting properties emerged~\cite{Zurek:2016j,Zurek:2018m}. Due to the sheer number of possible ternary chemical combinations and the cost of performing DFT-based CSP searches for them~\cite{Zurek:2020k}, some of the subsequent ternary searches were conducted using high-throughput methods. In such studies, once an optimal template is discovered, new structures are generated via atomic substitution of elements with similar chemistries in a given supercell~\cite{GENG2024101443,PhysRevLett.128.047001,PhysRevB.105.104511,lucrezi2022silico}. 
As a result, the structural diversity in the ternary dataset is somewhat reduced. However, in our study, special attention has been paid to ensure a dataset as heterogeneous as possible in terms of chemical composition and structural phases. The most characteristic structural patterns for the ternary compositions in our dataset are illustrated in Figure~\ref{FIG:structures}. 
Furthermore, we suggest a new classification scheme that is based upon the primary bonding mechanism of the important hydrogen atoms, labeling them as \emph{electropositive}, \emph{covalent} or \emph{interstitial}. 

%

\begin{figure*}
\includegraphics[width=\textwidth]{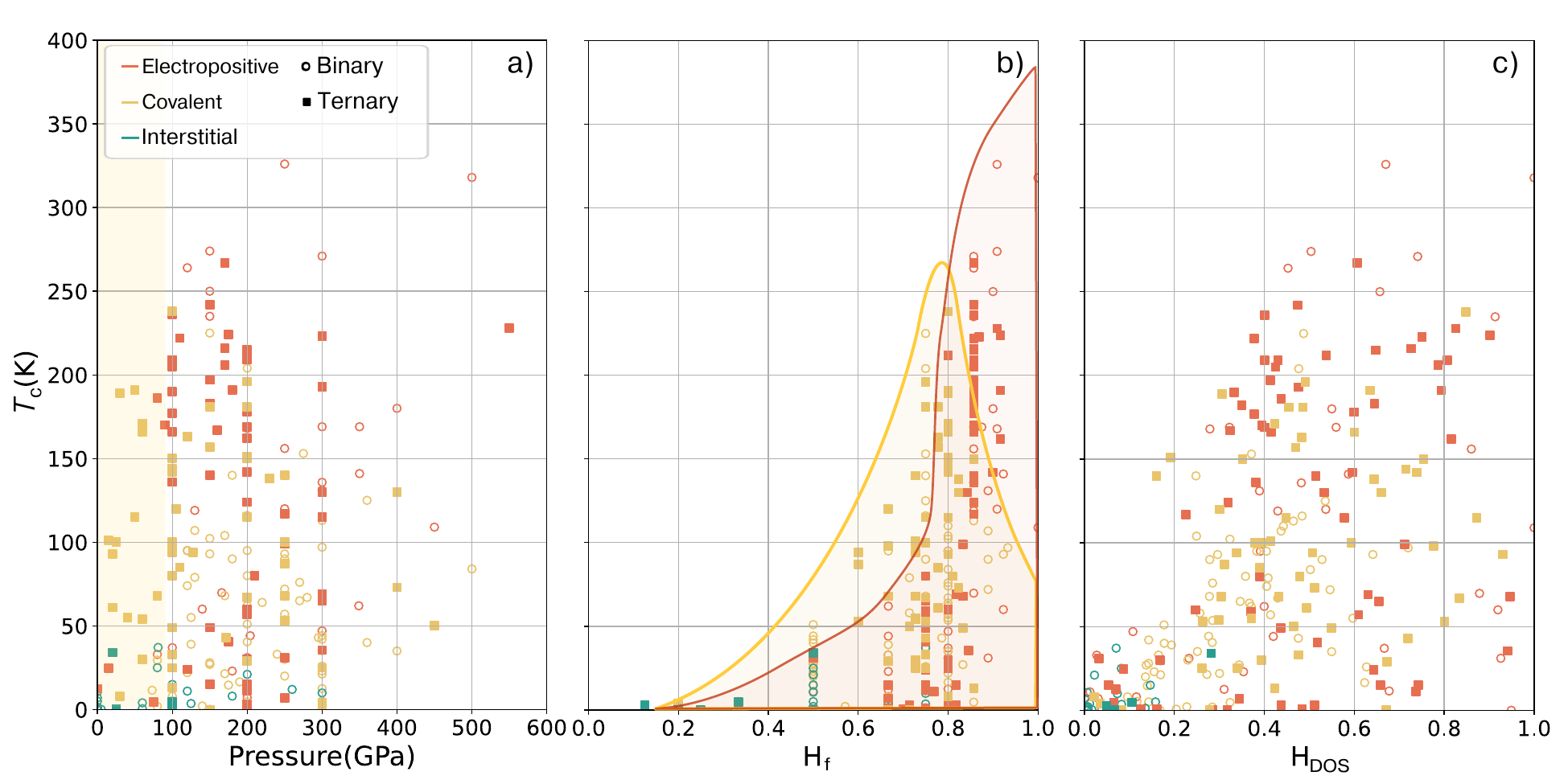}
\caption{\label{FIG:trends} \tc\ as a function of (a) pressure, (b) hydrogen fraction ({\hf}), and (c) the percentage of the density of states at the Fermi level projected onto the hydrogen atoms (H$_\text{DOS}$). Colors indicate the chemical classification of the systems: orange for \emph{electropositive}, yellow for \emph{covalent}, and green for \emph{interstitial} systems. Markers differentiate system types: empty circles for binary compounds and squares for ternary compounds. The shaded region in panel (a) highlights the range where the \tc\ values of \emph{covalent} systems are significantly higher compared to those of the other classes. The shaded areas in panel (b) represent the maximum {\tc} trends for both \emph{electropositive} and \emph{covalent} systems. }
\end{figure*}

The \emph{electropositive} hydrides are represented by compounds 95, 174, and 7 in Figure \ref{FIG:structures}.  \emph{Every} non-hydrogen atom in these compounds is a highly electropositive element whose 1~atm electronegativity is approximately lower than 1.6 on the Pauling scale (neglecting the changes in electronegativity under pressure~\cite{doi:10.1021/jacs.9b02634}), such as alkali metals, alkaline earth metals, and rare earth metals. Some early and post-transition metal elements also fall in this class; examples include Al (forming AlH$_3$~\cite{Goncharenko:2008a}), Zr (predicted to adopt $I4/mmm$ ZrH$_4$ and $I4/mmm$ ZrH$_6$ phases, which resemble those found for Group 2 polyhydrides~\cite{Abe:2018a}),  and Ta (forming TaH$_6$~\cite{doi:10.1021/acs.inorgchem.6b02822}). The bonding between the metal and hydrogen atoms is ionic, however for some metals, including Ca, Sr or Ba Kubas-like bonding between hydrogen and the metal d-states can occur~\cite{Zurek:2018b,Zurek:2020g}. The metal ions act as spacers as well as electron donors, and the predicted \tc s can span a very wide temperature range (Figure \ref{FIG:trends} (a)). 

To better understand why the \tc\ in this class of compounds varies so drastically, we consider the broad range of hydrogenic motifs present. Their emergence can be explained by considering the number of ``effectively added electrons'' (EAE) per H$_2$; if the number is significant, the H-H bond is weakened and eventually broken due to filling the H$_2$ $\sigma^*$ anti-bonding orbitals, but for smaller EAE values, the H-H bonds are simply stretched, making it possible for extended hydrogenic lattices to form~\cite{Wang:2012}. Importantly, the types of hydrogenic motifs present dictate the mechanism of metallization~\cite{Zurek:2016j,Zurek:2018m}. Phases that contain H$^-$ and H$_3^-$ units become metallic due to pressure-induced band-broadening, typically resulting in a low DOS at the Fermi level and a low \tc\ (e.g.\ MgSc$_3$H$_{12}$, \#174). Phases with H$_2^{\delta-}$ units, which possess slightly stretched H-H bonds, become metallic from electron donation into the anti-bonding H$_2$  $\sigma^*$ orbitals, resulting in a high DOS at the Fermi level and respectable \tc s (e.g.\ YMgH$_8$, \#7) . However, the {\tc}s of systems with even more complex molecular hydrogenic motifs (such as the H$_{10}^{\delta -}$ pentagraphene-like units~\cite{Zurek:2017j,Xie:2020a}), or 1D and 2D hydrogenic lattices are often found to be even higher, while those with 3D extended lattices are predicted to have the highest \tc s (e.g.\ LuCaH$_{12}$, \#95). 

Many of the \emph{electropositive} hydrides with more complex or highly dimensional hydrogenic motifs are highly symmetric, typically possessing a hydrogen fraction ({\hf}) of at least 0.8 (Figure \ref{FIG:trends}(b)). Without pressure they would be unstable, and the hydrogenic lattices would decompose into H$_2$ molecules and H$^-$ anions. As a result, no structure of this kind exhibits high \tc\ below 100~GPa (Figure \ref{FIG:trends}(a)). A large EAE is characteristic of a low hydrogen content, favouring the formation of hydridic, H$^-$, anions~\cite{Wang:2012}. While phases containing hydridic hydrogens remain stable at lower pressures and indeed are often the 1~atm ground state, their highest {\tc}s are significantly lower, and many may even be non-metallic. Systems with {\tc}s intermediate to these two extremes are represented by the MH$_4$ family, distinguished by the synthesis of its many members, with CaH$_4$ persisting on decompression to 60~GPa at room temperature~\cite{Zurek:2018b}. While CaH$_4$ can be viewed as $(\text{Ca}^{2+})(2\text{H}^-)+\text{H}_2$, and hence is non-metallic, adding one more electron to this system and achieving an (average) metal valence of +2.5 results in metallicity and superconductivity, as in YMgH$_8$. The factors responsible for the emergence of superconductivity in the $I4/mmm$ MH$_4$ structure type have been fully analyzed~\cite{Zurek:2020a}.

The second class of compounds we consider are \emph{interstitial} hydrides, where the non-hydrogen atoms are often transition metal elements with electronegativities approximately greater than 1.6~\cite{Zurek:2018d}. In their elemental form many of these metals assume simple lattices (e.g.\ $B1$ or $B2$, face-centered or body-centered cubic, and hexagonal close-packed), and hydrogen can easily fit into their empty octahedral or tetrahedral sites.  One member of this class of compounds is the predicted PdCuH$_2$ phase (\#186 in Figure \ref{FIG:structures}), where the PdCu lattice assumes the $B2$ structure type and hydrogen fills the octahedral holes~\cite{belli2024efficient}. Though these phases are often stable at ambient pressures, and many are metallic, the Fermi level is dominated by metal d-states, resulting in low {\tc}s (Figure \ref{FIG:trends}(a)), although recently an anharmonic \tc\ of 66~K was predicted for Cu$_4$H$_3$ at 1~atm~\cite{D5MH00177C}. In this class of compounds, H$_\text{f}\lesssim 0.5$; for higher concentrations, the hydrogen atoms can no longer be confined to the interstitial regions, and other motifs form. For example, NbH$_x$ ($0<x<0.7$) falls in the \emph{interstitial} class, exhibiting \tc\ up to 9.4~K, but compounds with $x=3.49-3.77$ would be classified as \emph{electropositive} with a measured \tc\ of 34~K at 133~GPa~\cite{doi:10.1021/jacs.4c15868,physRevB.88.184104}.


Finally, the class of \emph{covalent} hydrides remains the same as in the initial ELF-based manuscript~\cite{belli2021strong}. It includes compounds where (polar) covalent bonds are formed between hydrogen and a p-block element with Pauling electronegativity approximately greater than 1.6 (e.g.\ \#79 in Figure \ref{FIG:structures}), as well as ones containing complex anions that feature hydrogen bonded to a p-block or transition metal element in the presence of an electropositive countercation (e.g.\ \#179, 84 and 51 in Figure \ref{FIG:structures}).  When covalent bonds involving hydrogen and elements such as B, Si, P, S, and Se are present, along with a large DOS at the Fermi level, \tc\ can be significant. High-pressure often results in increasing coordination numbers, and p-block elements can adapt by making use of hypervalent bonding schemes~\cite{Grochala:2007a}. Indeed, the coordination environments in the examples that we illustrate from this class are unusually high. This includes octahedrally coordinated S and Se atoms in SeSH$_6$, 12-coordinate P atoms in LiPH$_6$, octahedrally coordinated B atoms in Li$_2$BH$_6$, or 8-fold coordinated P atoms in LaPH$_8$. And, in a few of these phases, hydrogen forms bridging bonds, with a bond order of $\sim$1/3~\cite{Zurek:2021i}, or bonds analogous to those in found in diborane~\cite{Zurek:2017c}. Though hypercoordination and multi-centered bond formation increase coordination numbers, geometric factors limit the maximum {\hf} in the highest \tc\ systems. Increasing hydrogen content results in the formation of H$_2$ molecules above H$_\text{f}=0.8$, resulting in a drop of the \tc\ (Figure \ref{FIG:trends} (b)). 

The interesting feature of this class of \emph{covalent} hydrides is that the barriers to breaking the bonds between the p-block element and hydrogen are typically larger than those required to sever the weak H-H bonds in the unusual hydrogenic motifs at times observed in the \emph{electropositive} family. Moreover, some members of this family feature well-known complex anions with unusual charges forced upon them by the composition of the countercations. Examples include K(BH$_4$)$_2$ (which has one electron too few to achieve a closed-shell BH$_4^-$)~\cite{Zurek:2023m} and Mg$_2$IrH$_6$ (which has one electron too many to achieve the closed-shell IrH$_6^{3-}$ configuration)~\cite{Zurek:2024i}. Both factors result in the dynamic stability of \emph{covalent} hydrides to much lower pressures than the ionic ones, sometimes even at atmospheric conditions. However, kinetic stability at such low pressures must be carefully evaluated since decomposition barriers may be low, and to the best of our knowledge, none of these phases have yet been synthesized, as effects such as configurational entropy~\cite{Hansen:2024}, may further penalize their formation. 

It should be noted that the chemical composition alone cannot be used to assign compounds to this class; the specific geometry needs to be considered to verify the presence of the hydrogen-containing heteronuclear (H-X) covalent bond. For example, H$_4$I, predicted to be stable at 100~GPa, contains H$_2$ units and a monoatomic iodine lattice \cite{Zurek:2015d}. And, in CaSH$_2$ ionic bonding between the negatively charged p-block element and the electropositive element is preferred over the formation of an X-H bond, resulting in a system that can be viewed as $(\text{Ca}^{2+})(\text{S}^{2-})+\text{H}_2$~\cite{Zurek:2020g}. 

\begin{figure*}
\includegraphics[width=\textwidth]{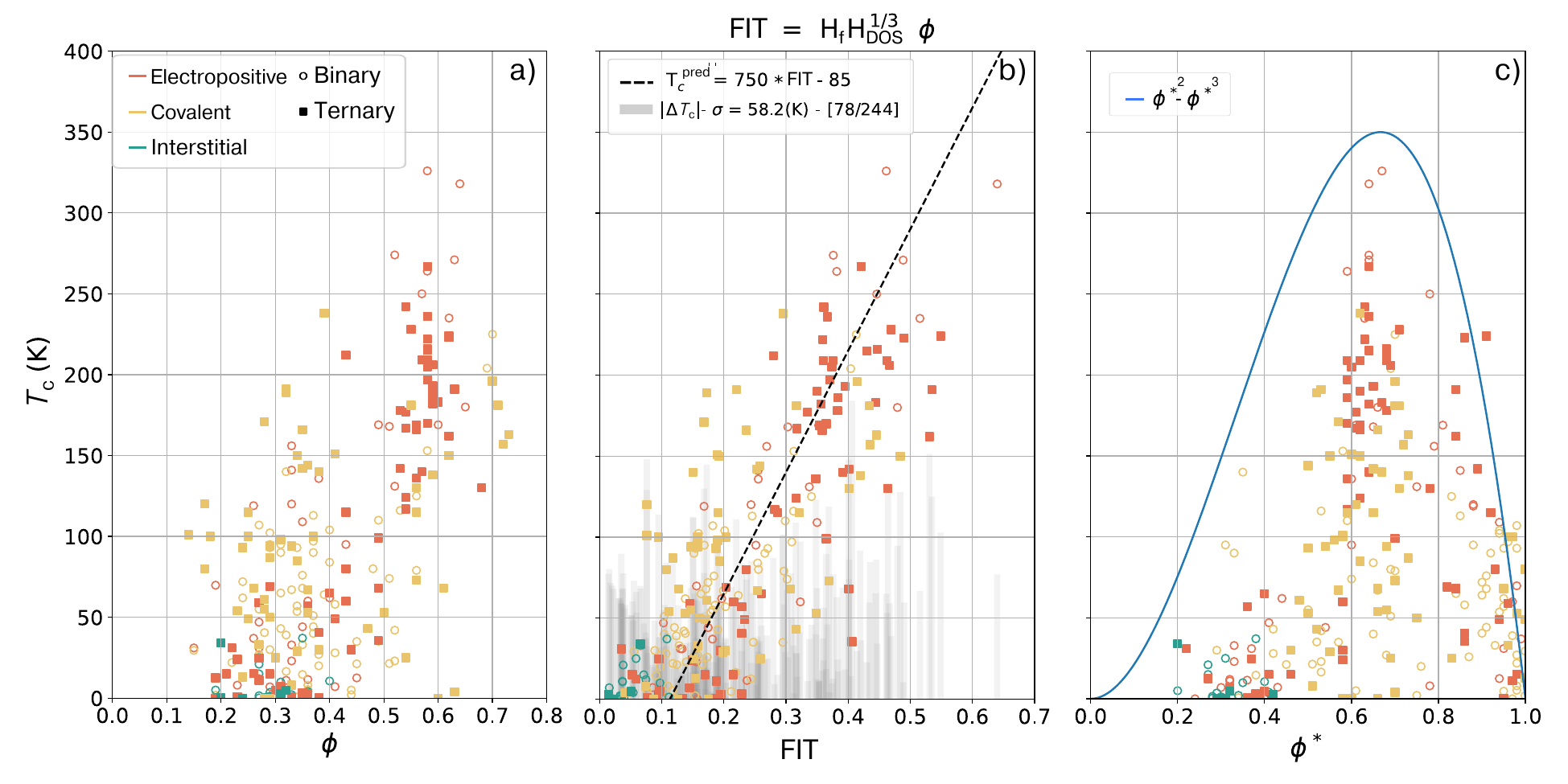}
\caption{\label{FIG:Netval} \tc\ as a  function of (a) the networking value $\phi$, (b) the standard fit for the \tc~\cite{belli2021strong}, and (c) the molecularity index $\phi^*$. Colors indicate the chemical classification of the systems: orange for \emph{electropositive}, yellow for \emph{covalent}, and green for \emph{interstitial} systems. Markers differentiate system types: empty circles for binary compounds and squares for ternary compounds. The dashed line in panel (b) represents the best fit proposed in Ref.\ \cite{belli2021strong}, while the gray lines represent the absolute value of the deviations of each point with respect to the proposed fit. The legend also provides the root mean square error ($\sigma$) and the fraction of compounds exhibiting a deviation greater than 60~K with respect to the fit in the square parenthesis ([78/244]) The blue line in panel (c) represents the $\phi^*$ term from Equation \ref{eq:betterFit}.}
\end{figure*}

\subsection{Networking and Superconductivity}

As alluded to above, high \tc\ values have been correlated with a large {\hf}. Can any other trends be found? Figure~\ref{FIG:trends}(b), illustrating the relationship between {\hf} and \tc\ for the extended set of binary and ternary hydrides used in this study, suggests different behavior for each of the three introduced classes of compounds. In \emph{electropositive} hydrides, a sharp decline in \tc\ is observed when {\hf} falls below 0.8 due to the insufficient the hydrogen concentration required to form the unusual motifs of weakly bonded hydrogen atoms responsible for high \tc\. In \emph{covalent} hydrides, the decrease in \tc\ is more gradual.  These systems are able to sustain superconductivity with lower hydrogen content as the vibrations of the H-X covalent bonds are involved in the electron-phonon coupling mechanism of superconductivity. The hydrogen fraction for \emph{interstitial} hydrides did not exceed 0.5 in any of the systems we considered, as hydrogen atoms fill lattice voids in this class.

Additionally, a large DOS at the Fermi level does not necessarily imply \tc\ will be high; instead, the fraction of this DOS that is attributed to hydrogen atoms, {\hdos}, is the key descriptor for favourable {\tc}. This characteristic derives from the atomic mass dependence of the phonon linewidth associated with the electron-phonon interaction. A high contribution of the hydrogen states to the DOS at the Fermi level increases the magnitude of the linewidth. A plot of the computed {\hdos} versus \tc\ (Figure~\ref{FIG:trends} (c)) shows similar behavior to that previously noted when only binary hydrides were considered~\cite{belli2021strong}. Generally speaking, \tc\ decreases with decreasing {\hdos}, though there is a large scatter in the data, potentially resulting from the hydrogenic contribution to the DOS at the Fermi level obtained via projection, whose results can be sensitive to the atomic radii chosen. Nonetheless, the trend appears to be independent of the group classification introduced, suggesting a more general validity of this descriptor. Additionally, it is noteworthy that below a value of 0.2, there is a sharp drop in \tc. This behavior was first observed in binary hydrides and continues to hold for ternary systems, yet it is unclear if this drop presents a physical meaning, or if it is an artifact due to lack of data or projection artifact onto {\hdos}.

It was also observed that the {\tc}s of hydrides correlate with the networking value ($\phi$), defined as the highest value of the ELF forming an iso-surface spanning the entire crystal in all three Cartesian directions~\cite{belli2021strong}. Figure~\ref{FIG:Netval}(a) shows the networking values for an extended dataset including both binary and ternary compounds. The inclusion of ternaries significantly increases the spread of the data. This broader distribution arises primarily from two types of structures: \emph{covalent} hydrides with low $\phi$ values but high {\tc}, such as those containing highly symmetric XH$_4$ or XH$_8$ units embedded in a metal matrix, e.g., K(BH$_4$)$_2$~\cite{PhysRevB.105.224107} and CaBH$_8$~\cite{PhysRevLett.128.047001}; and, at the other extreme, compounds with high $\phi$ values that form extended networks of bonded X-X atoms but exhibit low {\hdos} and {\hf}, such as CaBH~\cite{PhysRevB.102.014516}. By combining $\phi$ with {\hf} and {\hdos} a fit was proposed~\cite{belli2021strong} that could be used to estimate the \tc\ of binary hydrides (with an error of $\sim$60~K) as: 
\begin{equation} 
T_\text{c} = 750 \phi \mathrm{H}_\text{f} \sqrt[3]{\mathrm{H}_\text{DOS}} - 85.
\label{eq:Tc}
\end{equation}

To understand the applicability of Equation~\ref{eq:Tc} to ternary hydrides, we applied it to the systems considered and plotted the results in Figure~\ref{FIG:Netval}(b). Although the correlation still holds, with a root mean squared error, absolute mean error, and max variance for the full (binary-only) dataset being 59 (46), 45 (39), and 185 (60)~K, respectively, the spread of the data is much wider for the ternaries. Notably, almost 30\% of the data deviate from the reported value by more than 60~K. Furthermore, around values of 0.4 for the fit (Figure~\ref{FIG:Netval}(b)), the spread of the predictions with respect to the reported \tc\ values obtained from the literature is roughly 250~K, prohibiting an estimation of the \tc. 

Multiple factors likely contribute to this large deviation. First, \tc\ can be estimated with different computational models that differ in their accuracy. While the majority of the dataset contains \tc\ values calculated with the Allen-Dynes formula, for some compounds the \tc\ incorporated in the dataset was obtained through Migdal-Eliashberg theory (e.g.\ LaC$_2$H$_8$~\cite{https://doi.org/10.1002/pssr.202300043}). Although it has been noted that the Allen-Dynes and Migdal-Eliashberg frameworks can, in principle, both yield reliable results with a proper adjustment of the Coulomb pseudopotential, $\mu^*$, it is unlikely that these fine effects were considered in the current literature.

Additionally, some of the reported compounds are not kinetically stable. For a subset of hydrides, we assessed the resilience to overcoming barriers through molecular dynamics simulations at 50~K and at fixed pressures. During these runs some of the compounds dissociated or distorted. A few examples are PrH$_9$ (100~GPa)~\cite{PhysRevLett.119.107001}, whose cell transitioned from cubic to rhombohedral, LaBH$_8$ (100~GPa) and CaBeH$_{8}$ (210~GPa)~\cite{PhysRevB.104.L020511,PhysRevB.106.134509,PhysRevLett.128.047001}, where the XH$_8$ molecules dissociated into XH$_6$ and H$_2$,  and CeH$_9$ (50~GPa)~\cite{PhysRevLett.119.107001}, which underwent a cell distortion. In total, 39 randomly selected compounds were investigated through molecular dynamics simulations, and only 19 were found to maintain their structure. In this regard we expect that just 50\% of the dataset included is reliable, corresponding to local minima with sufficiently high barriers.

While the networking value is related to the electronic contribution to the \tc, it is not able to incorporate the effects of the atomic displacements, which also play a key role in determining superconducting properties. The vibrational contribution is one of the reasons why two isomorphic structures, differing only in their atomic constituents, can have drastically different \tc s,  and why modifications in the \tc\ by up to 50~K can be expected via the inclusion of anharmonicity, as in the case of LaBH$_8$~\cite{PhysRevB.106.134509}. An attempt to incorporate the effect of the atomic displacements (vibrational frequencies) on the \tc\ considered the average charge density per mass (Section S3). Although a correlation between \tc\ and this quantity was found, it did not present a sharp enough resolution to be able to suggest predictive trends.

\begin{figure}
\includegraphics[width=\columnwidth]{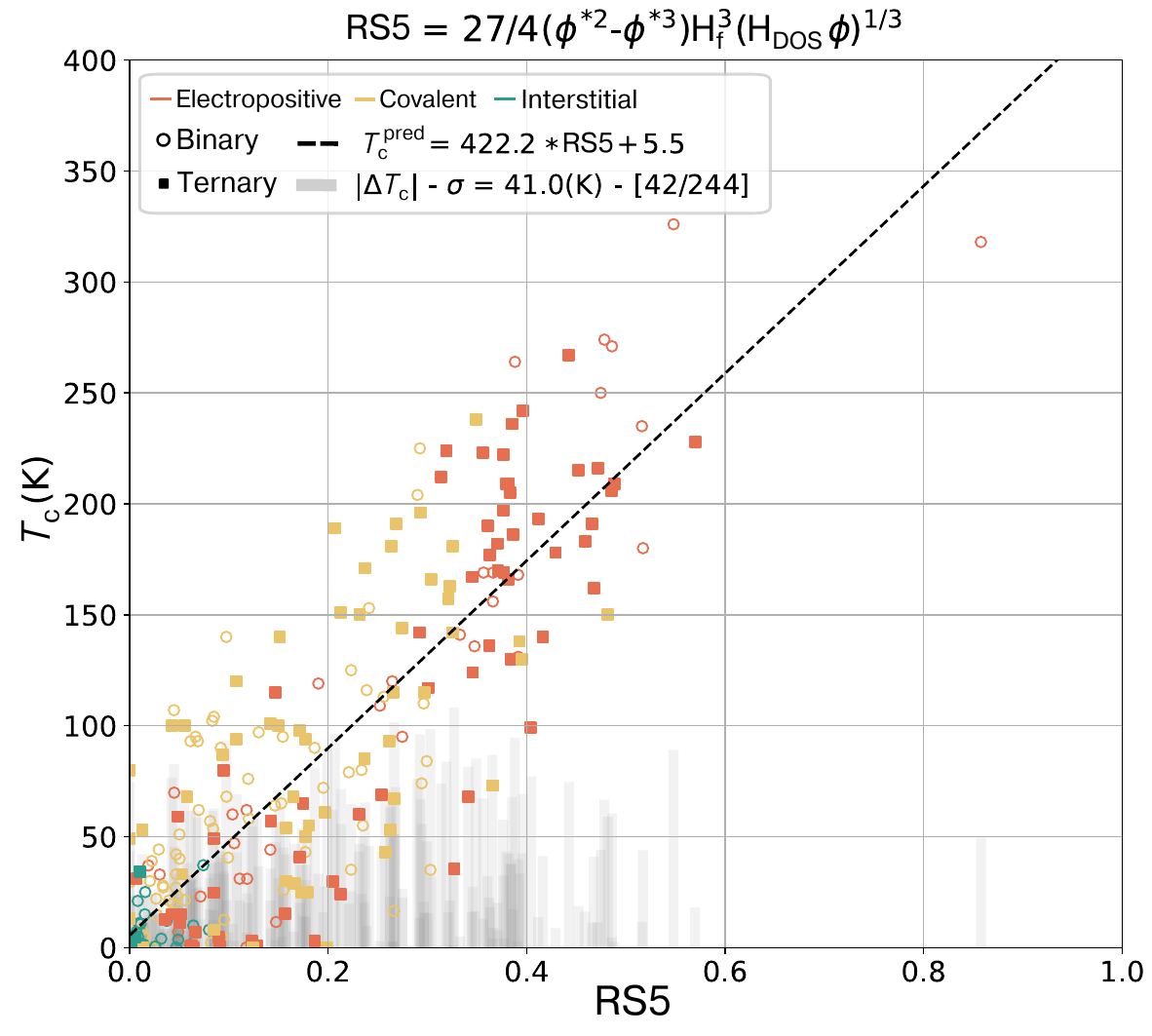}
\caption{\label{FIG:Newfit}  \tc\ values as a function of the new proposed fit for the \tc. Colors indicate the chemical classification of the systems: orange for \emph{electropositive}, yellow for \emph{covalent}, and green for \emph{interstitial} systems. Markers differentiate system types: empty circles for binary compounds and squares for ternary compounds.  The dashed line is the best fit, while the gray lines represent the absolute value of the deviations of each point with respect to the proposed fit. The legend also provides the root mean square error ($\sigma$) and the fraction of compounds exhibiting a deviation greater than 60~K with respect to the fit in the square parenthesis ([42/244]).}
\end{figure}

Di Mauro et al.\ argued that the networking value is not able to capture small local structural variations, such as symmetry breaking distortions, which can drastically affect \tc~\cite{PhysRevB.110.174515}. To overcome this problem the authors introduced the molecularity index ($\phi^*$), defined as the highest value of the ELF at which two hydrogen atoms become connected. The molecularity index reflects the emergence of the strongest interaction relative to the formation of the network, while the networking value captures the overall development of the network itself. When interactions are inhomogeneous, the molecularity index and the networking value diverge, which leads to a reduction in \tc. 

Figure~\ref{FIG:Netval}(c) reports the \tc\ versus the molecularity index for both binary and ternary hydrides. The same trends are observed for both sets, suggesting consistency.  Di Mauro et al.\ proposed a series of fits, some that employed the molecularity index, and identified the range $\phi^* \in [0.45, 0.8]$ as more likely to host compounds with higher \tc~\cite{PhysRevB.110.174515}. We applied these same fits to our extended dataset (Figure S3 and S4) and found that only one of them retained correlation with the extended dataset (Figure S4(d)). This fit turns out to be a slight modification of the networking value fit provided in Equation \ref{eq:Tc}.

In the current study we improved the fit for \tc\ by using symbolic regression with the following physically motivated input variables:  {\hf}, {\hdos}, $\phi^*$, and $\phi$. To prevent overfitting, the dataset was divided into a training set and a testing set containing, respectively, 70\% and 30\% of the data. The resulting fit is shown in Figure \ref{FIG:Newfit} and has the following functional form:
\begin{equation}
    T_{\mathrm{c}} = 422.2 \times \frac{27}{4}
    (\phi^{*2}-\phi^{*3})\mathrm{H}_\mathrm{f}^3(\phi \mathrm{H}_{\mathrm{DOS}})^{\frac{1}{3}} +5.5.
    \label{eq:betterFit}
\end{equation}
The new fit yields a root mean squared error, absolute mean error, and max variance of  41, 31, and 108~K, respectively, and halves the amount of structures whose predicted {\tc}s differ by more than 60~K from the DFT literature values. Additionally, the new fit is able to resolve the zones between 0.1 to 0.4 where previously there was a wide spread in the predicted and DFT results. While the fit uses the same \hf, \hdos\ and $\phi$  variables employed in Equation \ref{eq:Tc}, the main correction arises from the $\phi^{*2}-\phi^{*3}$ term introduced by the symbolic regression. This term reflects the distribution of the maximum values of \tc\ as a function of $\phi^*$  (Figure \ref{FIG:Netval} (c)), which peaks at $\phi^*=0.68$. Furthermore, the term decreases the predicted \tc\ for systems that contain molecular hydrogen, where the network of interactions appears at lower ELF values than the ELF values associated with the molecular hydrogen bonds. These ELF features suggest an inhomogeneous set of iterations between the hydrogens, which will push the states associated with molecular hydrogen away from the Fermi level, therefore reducing \tc.

The new fit proposed herein significantly enhances the accuracy of the \tc\ predictions for an extended dataset of binary and ternary hydrides, and offers improved transferability to more complex multinary systems. With the higher resolution and transferability, the fit could possibly be integrated with multi-objective CSP tools to bias searches for hydrides toward phases with higher {\tc}s, or in liaison with high-throughput screenings prior to performing more refined \tc\ calculations.

\section{Conclusions}

An extended dataset of predicted hydrogen-based superconductors and their superconducting critical temperatures, \tc s, is collected and curated to remove unreliable data points. As compared to previously published datasets, it contains approximately the same number of binary and ternary hydrides, and care has been taken to incorporate a broad range of structural motifs. Furthermore, to facilitate further research, the dataset has been made publicly available. The hydrides comprising this dataset are classified based on the dominant bonding mechanism for the hydrogen atoms, leading to the identification of three distinct categories: \emph{electropositive}, \emph{interstitial}, and \emph{covalent} hydrides.

Both \emph{electropositive} and \emph{covalent} hydrides are capable of behaving as high-temperature superconductors. The main distinction between the two classes is the non hydrogen atoms electronegativity, that for the \emph{electropositive} class is approximately lower than 1.6 on the Pauling scale. In \emph{electropositive} hydrides high \tc s are associated with a large fraction of hydrogen atoms (H$_\text{f}>0.8$), which results in the formation of weakly bonded extended hydrogenic lattices. However, these compounds typically require pressures exceeding 100~GPa for dynamic stability.  In contrast, \emph{covalent} hydrides can attain moderate to high \tc\ values with a lower hydrogen content and at significantly reduced pressures, owing to the strong bonds formed between hydrogen and transition metal or p-block elements.

Additionally, this study analyzes the ability of descriptors based upon the networking value, $\phi$, to predict \tc. While the original fit proposed in Ref.\ ~\cite{belli2021strong} continues to capture general trends, the increasing structural and electronic complexity of the compounds comprising the extended dataset reduces its predictive power. To address this limitation, symbolic regression is employed to obtain a new fit, incorporating the molecularity index proposed by Di Mauro et al.~\cite{PhysRevB.110.174515} as an additional variable. This new model provides a substantially improved estimate of \tc\ across all hydride types, achieving a root mean squared error of $\sim$40~K. The enhanced accuracy of the fit is attributed to the broader and more heterogeneous nature of the training dataset, which enables better generalization across diverse compounds and is expected to be more resilient for the prediction of \tc\ within of hydrides. We expect the proposed fit to be useful in future computational studies, especially in screening complex hydride-based structures for promising superconducting properties prior to performing expensive first-principles calculations.

\section{Methods}
Our analysis was carried out using the 127 hydrides collected by Belli et al.~\cite{belli2021strong} with the addition of 186 ternary hydrides obtained from the literature~\cite{PhysRevB.107.134509,https://doi.org/10.1002/adts.202100364,PhysRevB.104.L020511,PhysRevB.106.134509,PhysRevB.105.104511,PhysRevB.105.224107,PhysRevB.102.184103,PhysRevB.101.134504,PhysRevB.102.014516,PhysRevB.106.014521,PhysRevLett.128.047001,PhysRevB.106.024519,PhysRevB.100.184502,PhysRevB.104.224510,PhysRevB.104.134501,PhysRevMaterials.1.074803,doi:10.34133/2022/9784309,Tian_2015,Sun_2022,D2TC04029H,D2CP00059H,D2TC02842E,shi2021prediction,D1CP03896F,D1CP04963A,D1CP02781F,doi:10.1021/acs.jpclett.9b03856,D2CP05850B,tsuppayakorn2022stabilizing,doi:10.1021/acs.jpcc.1c08743,SAHOO2023111193,shao2019ternary,https://doi.org/10.1002/pssr.202300043,https://doi.org/10.1002/er.8705,PhysRevB.96.144518,C7CP05267G,doi:10.1021/acs.jpcc.0c09447,10.1063/5.0076728}. Each ternary compound was initially relaxed at the pressure corresponding to its maximum reported {\tc}, and only the structures that did not exhibit significant distortions affecting symmetry, cell shape, or causing substantial renormalization of atomic distances were retained. A subset of structures was analyzed to assess kinetic stability, which depends on the barrier heights that protect the structure from decomposition. This subset of 38 compounds consisted of 17 ternaries and 21 binaries. While a number of computational techniques could be used to determine kinetic stability, we have chosen to perform \emph{ab initio} molecular dynamics (MD) simulations at specific temperatures. Compounds that underwent a significant structural distortion within the first 2~ps of the MD run were discarded from the dataset. While global CSP searches rarely find kinetically unstable species, high-throughput techniques are more prone to identify dynamically stable structures that might decompose at finite temperatures~\cite{Zurek:2023m}. The final dataset used in our analysis, after discarding the distorted and kinetically unstable structures, consisted of 119 binary and 125 ternary compounds, for a total of 244 structures (78.6\% of the original dataset).

All of the DFT calculations were performed using the plane-wave {\sc Quantum ESPRESSO} (QE) package~\cite{Giannozzi1,Giannozzi2}. The exchange-correlation potential was approximated using the Perdew-Burke-Ernzerhof (PBE) parameterization \cite{GGA-PBE} within the projector-augmented wave (PAW) method. The plane-wave cutoff energy was set to 90~Ry, while 900~Ry was used for the charge density. Brillouin zone integrations were conducted using the Methfessel-Paxton smearing method~\cite{Methfessel} with a broadening of 0.02 Ry. The \textbf{k}-point grids were chosen with densities of $2\pi\times 0.036$ \AA$^{-1}$ for self-consistent calculations, and $2\pi \times0.018$ \AA$^{-1}$ for non-self-consistent calculations. Electronic properties, including the ELF, DOS, and charge distributions, were computed using QE post-processing tools based on the non-self-consistent calculations, and a Bader analysis \cite{bader1990atoms} was used to obtain the charge and volume per atom. The calculation of the ELF critical points was performed through the {\sc Critic2} program~\cite{OTERODELAROZA20141007,OTERODELAROZA2009157}, and subsequently the networking value and the molecularity index were obtained using {\sc TcEstime}~\cite{D4SC04465G} and double-checked by hand. 

The MD simulations were performed using the Vienna \emph{ab initio} Simulation Package ({\textsc VASP})~\cite{hafner2008ab}, with the PBE functional~\cite{GGA-PBE} and an energy cut-off of 800~eV, treating the core electrons with the projector augmented wave (PAW) method~\cite{PhysRevB.50.17953}. The simulations were performed on a $\Gamma$-only \textbf{k}-grid at 50~K for 10~ps in the $NPT$ ensemble, with the Langevin thermostat, and using a 3$\times$3$\times$3 supercell expanded from the primitive cell after relaxing the structure to the chosen pressure through QE. 

For the symbolic regression analysis, the Sure Independent Screening and Sparsifying Operator (SISSO)~\cite{PhysRevMaterials.2.083802} was employed to model both binary and ternary compounds. The dataset was split into a training set (70\% of the data) and a test set (30\%). The symbolic feature space was constructed using the following mathematical operators: $+, -, *, \cdot/\cdot, \cdot^{-1}, \cdot^{2}, \cdot^{3}, \cdot^{6}, \sqrt{\cdot}, \sqrt[3]{\cdot}$, and $|\cdot|$. The optimal descriptor was determined by minimizing the maximum absolute error.\\

\section{Acknowledgements}
Funding for this research is provided by the National Science Foundation, under award DMR-2136038 and the US Department of Energy, Office of Science, Fusion Energy Sciences under award DE-SC0020340, entitled \emph{High Energy Density Quantum Matter}. Calculations were performed at the Center for Computational Research at the State University of New York at Buffalo (https://hdl.handle.net/10477/79221). \\

\section{Author Contributions}
The project was conceived and supervised by E.Z. F.B.\ retrieved the data and performed the calculations and analysis, with assistance from S.T. F.B.\ wrote the manuscript and S.T.\ contributed to the writing. E.Z.\ edited the manuscript.\\

\section{Data Availability}
The data supporting this study's findings are available on figshare. The structures for the ternary systems can be accessed at \href{https://doi.org/10.6084/m9.figshare.28856945.v1}{https://doi.org/10.6084/m9.figshare.28856945.v1}, the data used in the analysis for all systems can be found at \href{https://doi.org/10.6084/m9.figshare.28856786}{https://doi.org/10.6084/m9.figshare.28856786}.\\

\section{Competing Interests}
The authors declare no competing interests. \\









\bibliography{bibliography}

\end{document}